\documentclass[a4paper,11pt]{article}

\usepackage{amsmath, amsfonts, url}
\usepackage{authblk}
\usepackage{graphicx}
\graphicspath{{fig/}}
\usepackage{a4wide}
\usepackage[parfill]{parskip}
\usepackage{geometry}
\usepackage{pgf,tikz}
\usetikzlibrary{arrows}
\usepackage{rotating}
\usepackage{verbatim}
\usepackage[parfill]{parskip}
\usepackage{natbib}

\newcommand{\ie}{{\emph{i.e.}}}

\title{Pachinko Prediction: A Bayesian method for event prediction from social media data}

\author[1,2]{Jonathan Tuke}
\author[1]{Andrew Nguyen}
\author[1,2]{Mehwish Nasim}
\author[3]{Drew Mellor}
\author[3]{Asanga Wickramasinghe}
\author[1,2]{Nigel Bean}
\author[1,2,4]{Lewis Mitchell}

\affil[1]{School of Mathematical Sciences, University of Adelaide, SA 5005, Australia}
\affil[2]{ARC Centre of Excellence for Mathematical and Statistical Frontiers (ACEMS)}
\affil[4]{Data to Decisions Cooperative Research Centre (D2D CRC), Kent Town, SA 5067, Australia}
\affil[4]{D2D CRC stream lead}

\providecommand{\keywords}[1]
{
  \small	
  \textbf{\textit{Keywords---}} #1
}

\date{}

\begin{document}

\maketitle


\begin{abstract}
The combination of large open data sources with machine learning approaches presents a potentially powerful way to predict events such as protest or social unrest.
However, accounting for uncertainty in such models, particularly when using diverse, unstructured datasets such as social media,
is essential to guarantee the appropriate use of such methods.
Here we develop a Bayesian method for predicting social unrest events in Australia using social media data.
This method uses machine learning methods to classify individual postings to social media as being relevant, and an empirical Bayesian approach to calculate posterior event probabilities.
We use the method to predict events in Australian cities over a period in 2017/18.

\hspace{10pt}

\keywords{Bayesian statistics, social unrest, machine learning, prediction}
\end{abstract}

\section{Introduction}


Developing automated methods to give advance warning of large gatherings of people,
such as protests and social unrest events,
are of interest to government agencies worldwide.
With such events often being organised over online social media platforms,
there exists the possibility to provide prior warning of large events solely through monitoring online data streams.
Researchers have used open online data sources such as Twitter \citep{Borge-Holthoefer2015a,Agarwal2016}, 
Facebook,
Tumblr \citep{Xu2014},
and Flickr \citep{Alanyali2015}
to characterise information propagation processes around protests,
and have deployed machine learning methods on social media
as well as blogs, news sources, and the dark web \citep{Korkmaz2016} to predict civil unrest events.
Twitter data in particular has been used broadly to monitor diverse large-scale trends such as stock behaviour \citep{Bollen2011}, 
public opinion polling around issues like climate change \citep{Cody2015}, 
and health characteristics \citep{Alajajian2017}.
Recent studies have focussed on Twitter's role in particular in mobilisation and discourse around protest action in the United States \citep{Theocharis2015,Gallagher2018}.

Of particular note, and a catalyst for much of the research in this area, the US IARPA (Intelligence Advanced Research Projects Activity) OSI (Open Source Indicators) program\footnote{https://www.iarpa.gov/index.php/research-programs/osi} was started in 2011 to provide analysts with advance warning of civil unrest events in South America based on open data gathered from social media and other sources.
This contest provided a ``gold standard record'' of event timings and locations to teams in order for them to develop (usually supervised) machine learning approaches to predict future events from OSI.
At that time, sophisticated learning approaches employing the fusion of multiple model outputs into a single prediction were found to be particularly effective, with the EMBERS (Early Model Based Event Recognition using Surrogates) model \citep{Ramakrishnan2014} ultimately providing the best predictions according to the evaluation metrics set for the contest.
A key distinguishing feature of this model compared with other machine learning approaches was the use of multiple models combined via fusion \citep{Hoegh2015}, along with a novel method for suppressing spurious model outputs.

One general characteristic of the models forming inputs to EMBERS, as well as of many other models in the IARPA OSI program,
was that they produced binary predictions of future days as having an event or not.
Underlying each model in EMBERS was a ``hard'' (binary) classification of future day-location pairs into event/non-event,
either from a rule-based scheme, e.g., the so-called ``Planned Protest" model \citep{Muthiah2015}, or by converting probabilities from a GLM such as logistic regression to a binary output, e.g., the ``Volume-based'' \citep{Korkmaz2016} or ``Cascade'' models \citep{Cadena2015}.
This was likely guided by the evaluation methodology set out for the OSI program by IARPA.
Teams' model predictions were given an overall ``quality score'' comprising scores for date, location, event type, and population group,
and each score only took ``hard'' classifications as inputs.
For example, the ``date score'' DS for the competition was defined to be $DS = 1- \min\left(|\text{Event date} - \text{Predicted event date}|,7\right)/7$,
with no ability to account for a model's confidence in or uncertainty about the predicted event date.

In the context of the challenge this was likely a reasonable choice,
however it leaves open an important question: how confident was each model (or fusion of models) in each prediction?
If all models in the OSI program tended to lend only slightly higher weight to particular event days, then in the presence of randomness the outcomes of the program could be largely a matter of chance.
While this may not have been an issue with the OSI program (although it is difficult to know without access to the models),
it nonetheless opens avenues for further investigation into prediction from unstructured, ``noisy'' sources such as social media.
In particular, for analysts potentially using the outputs of forecasting systems such as EMBERS it is surely of interest to have access to a measure of the system's confidence around a given prediction, rather than a simple binary classification.

Thinking further from the perspective of a potential end-user of an automated forecasting system gives other desirable characteristics.
Ideally one would be able to disentangle the components of a prediction coming from different data sources,
or different models in a fusion framework.
For example, for a given predicted event day, knowing that the prediction was because that day is a typical protest day (e.g., Labor Day, or Australia Day in Australia) is very different to if the prediction was due to a sudden spike in social media activity which appeared suddenly.
Such a framework would potentially impart greater understanding, rather than mere predictions, of upcoming events.

In this paper, we develop a framework for predicting social unrest events from social media which incorporates these features.
We adopt an empirical Bayesian approach,
where a prior belief about the days and locations likely to see events is made explicit,
and then evidence (in the form of social media postings) is used to update this prior.
We focus on Twitter as an example of social media, however our methodology could be transferred to other social media platforms as well. 
Each individual tweet will form our smallest unit of observation. As well as giving a probabilistic interpretation of predictions and enabling us to disentangle different components of a prediction (through the prior and likelihood),
this framework empowers a simple conceptual understanding of the model.
As our algorithm involves sorting pieces of evidence into bins for different days and locations,
we adopt an analogy of coloured marbles being sorted into jars,
and hence the name ``Pachinko Prediction''.

The structure of the paper is as follows.
In Section \ref{sec:datasets}, we describe the ``Gold Standard Record'' of events in Australia over 2017/18 as well as the Twitter dataset used for training the model.
In Section \ref{sec:bayes_model}, we describe the Pachinko Prediction model framework and detail the underlying empirical Bayesian method.
Section \ref{sec:results} evaluates how the model performs at predicting events in Australian cities, 
and explores some qualitative features of the model.
We conclude with a discussion in Section \ref{sec:discussion}.

\section{Datasets}
\label{sec:datasets}
\subsection{Gold standard record}\label{sec:GSR}

A key component of our method, or any social unrest prediction effort such as the OSI program, is a ``Gold Standard Record'' (GSR) of historical events in a region.
Our project focuses on protests in Australia, 
therefore we use a custom-built GSR dataset specific to this region.
The GSR dataset is generated by analysts who were employed on a casual basis to read major news websites daily and record articles on any civil unrest events within Australia. 
Each of these events are recorded with additional attributes such as event name, location, time, and whether the event was violent or non-violent. 
The dataset included manual verification to correct any errors in the gathering.

\begin{sidewaysfigure}[htbp]
	\centering
	\includegraphics[width = 0.8\textwidth]{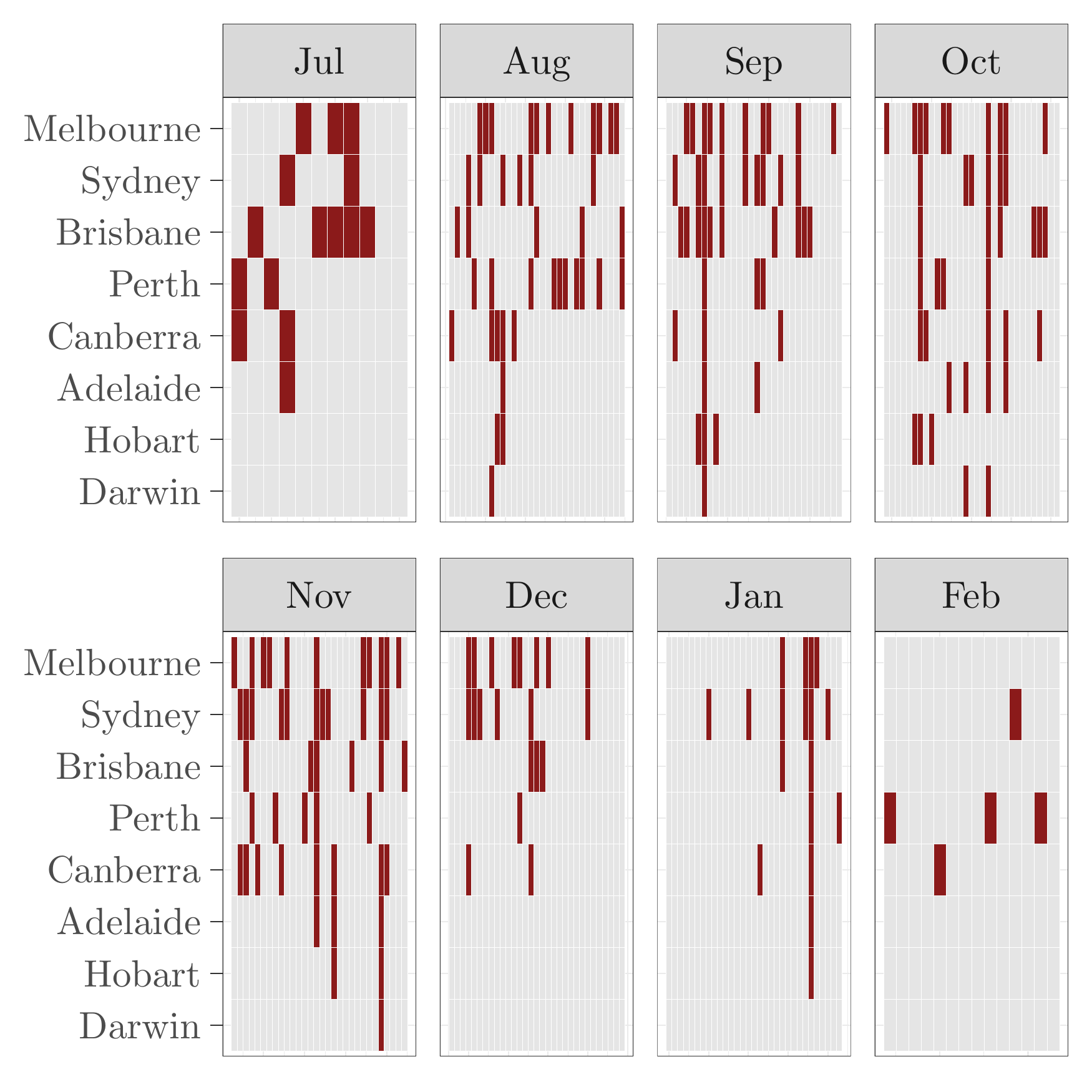}
	\caption{Time plot of GSR dataset. Red tiles indicate an event on that day/city combination.\label{fig:time_plot}}
\end{sidewaysfigure}

For this study we consider all the events for Adelaide, Brisbane, Canberra, Darwin, Hobart, Melbourne, Perth, and Sydney between the dates 
21 July 2017 to 14 February 2018 inclusively.
Figure~\ref{fig:time_plot} gives a tile-plot of each day for each city, with red tiles indicating events. We have ordered the cities by decreasing total number of events, and observe that Melbourne and Sydney -- the two largest cities in Australia -- had more observed events over the period.

Initially, we aim to predict whether an event occurs on a given day or not. 
We define the random variable $E_{ij}$ to be a Bernoulli random variable, equal to 1 if an event occurs on day $i$ in location $j$, and 0 if it does not. 

\begin{figure}
\centering
\includegraphics[width = \textwidth]{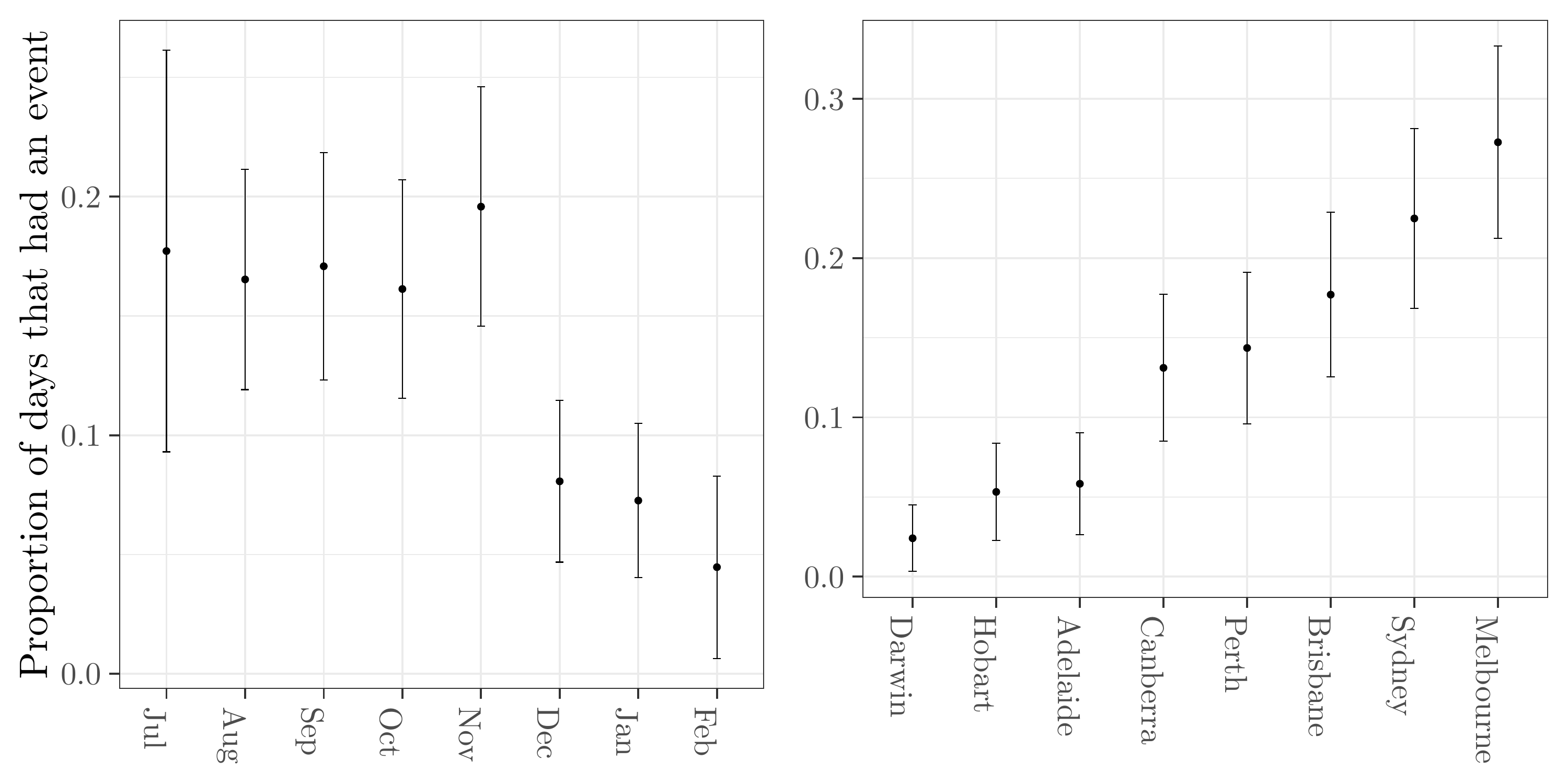}
\caption{Proportion of days that had an event for each month (left) and each city (right). The error-bars are 95\% confidence intervals for the proportion.}
\label{fig:city_month_prop}
\end{figure}

We first tested if there was a statistically significant difference in the proportion of days that had an event for the predictors: month, weekday, and city. 
We found a significant association between month and events (Chi-squared test, $\chi^2 = 36.808$, P-value $= 5.101 \times 10^{-6}$), 
and also city and events (Chi-squared test, $\chi^2 = 96.232$, P-value $< 2.2 \times 10^{-16}$). 
We did not find a significant association between weekday and events.
To illustrate, Figure~\ref{fig:city_month_prop}  shows the proportion of days having an event for each month (left), and the proportion of days having events for each city (right). 
There is a significant decrease in the proportion of days having events in December, January, and February compared to the other months. 
One possible explanation is that these are the summer months in Australia, 
and hot weather may decrease the number of events. 
Larger cities -- Melbourne, Sydney, and Brisbane -- have a significantly increased proportion of event days. 
This is intuitive, due to the larger pool of people who might be involved in protest events in these cities.

\subsection{Twitter}

Our Twitter data was collected using the public API\footnote{\url{https://developer.twitter.com/en/docs.html}} between 21 July 2017 to 14 February 2018 inclusive.
Roughly 50 million tweets were ingested per month, totalling approximately 350 million tweets.
We applied three filters on \emph{location}, \emph{temporal}, and \emph{relevance} characteristics, to reduce this collection to the (much smaller) final dataset we used for making predictions.
The \emph{location} filter aimed to target tweets relevant to Australian capital cities, by including all tweets matching any of the following criteria:
\begin{itemize}
    \item the ``Location'' field of the tweet's bio information contains the name of an Australian capital city, or 
    \item the tweet is geolocated to within a 25 mile radius of the centre of each Australian capital city, or
    \item there is a mention of an Australian capital city within the tweet body.
\end{itemize}


The \emph{temporal} filter selected only \emph{future-referencing} tweets by scanning the body of the tweet and resolving any time references mentioned. 
For example, a tweet published on the 2nd of January 2018 containing the sentence ``Let's protest tomorrow at the University of Melbourne'' would be resolved to the 3rd of January 2018, by resolving the ``tomorrow'' in the text.
We made this choice because our primary interest is in prediction, and so we wish to utilise only tweets that reference events in the future, rather than events that have already occurred (e.g., news reports).
This step is fundamental to our approach;
experimentation with purely ``volume-based'' models using tweets resolved to the date they were authored (e.g., \cite{Korkmaz2016}) produced poor results, with news reports of previous events swamping any potential signal.
We used Stanford NLP's SUTime \citep{Chang2012} to identify temporal mentions
along with HeidelTime \citep{Stroetgen2013} for processing multi-lingual tweets.
We applied the \emph{location} and \emph{temporal} filters simultaneously;
doing so left us with 51259 tweets.
The numbers of tweets from each location are given in Table \ref{tab:tweet_nums}.

\begin{table}[htbp]
\centering
\begin{tabular}{l|r}
  \hline
City & Tweets \\ 
  \hline
Adelaide & 2212 \\ 
  Brisbane & 3327 \\ 
  Canberra & 2370 \\ 
  Darwin & 270 \\ 
  Hobart & 565 \\ 
  Melbourne & 19980 \\ 
  Perth & 2602 \\ 
  Sydney & 19933 \\ 
   \hline
\end{tabular}
\caption{Number of future-referencing tweets collected for each location in our dataset.} 
\label{tab:tweet_nums}
\end{table}

Our \emph{relevance} filter was a custom civil unrest classifier used to select only tweets of interest.
This classifier was created as follows.
We manually labelled a random sample of 7898 tweets, containing 1504 positive examples linked to GSR events, and 6394 negative examples. 
This training set is made available along with this paper \citep{Mitchell2018a}.

We then tokenised tweets using the \texttt{CountVectorizer} in Python scikit-learn \citep{scikit-learn} to create 1- and 2-grams from the text. 
These tokens formed the features for the classifier.
Model selection was performed between four models: Gaussian and Bernoulli Naive Bayes, and a linear SVM with $L_1$ and $L_2$ penalty functions, on the basis of highest F1 score using 5-fold cross-validation.
The best-performing model selected was the linear SVM with $L_2$ penalty, having an F1 score of 0.94.
All models were implemented using scikit-learn with default parameters.

Applying this classifier to the data left us with 51,259 tweets.
We remark that while this is a relatively small dataset compared to typical modern ``big data'' approaches,
it is the result of an extensive filtering procedure designed to leave us mostly with informative tweets regarding the events of interest.
This Twitter dataset formed the input training data for our model, which we describe in the next section.



\section{Method}
\label{sec:bayes_model}

\begin{figure}
\includegraphics[width = \textwidth]{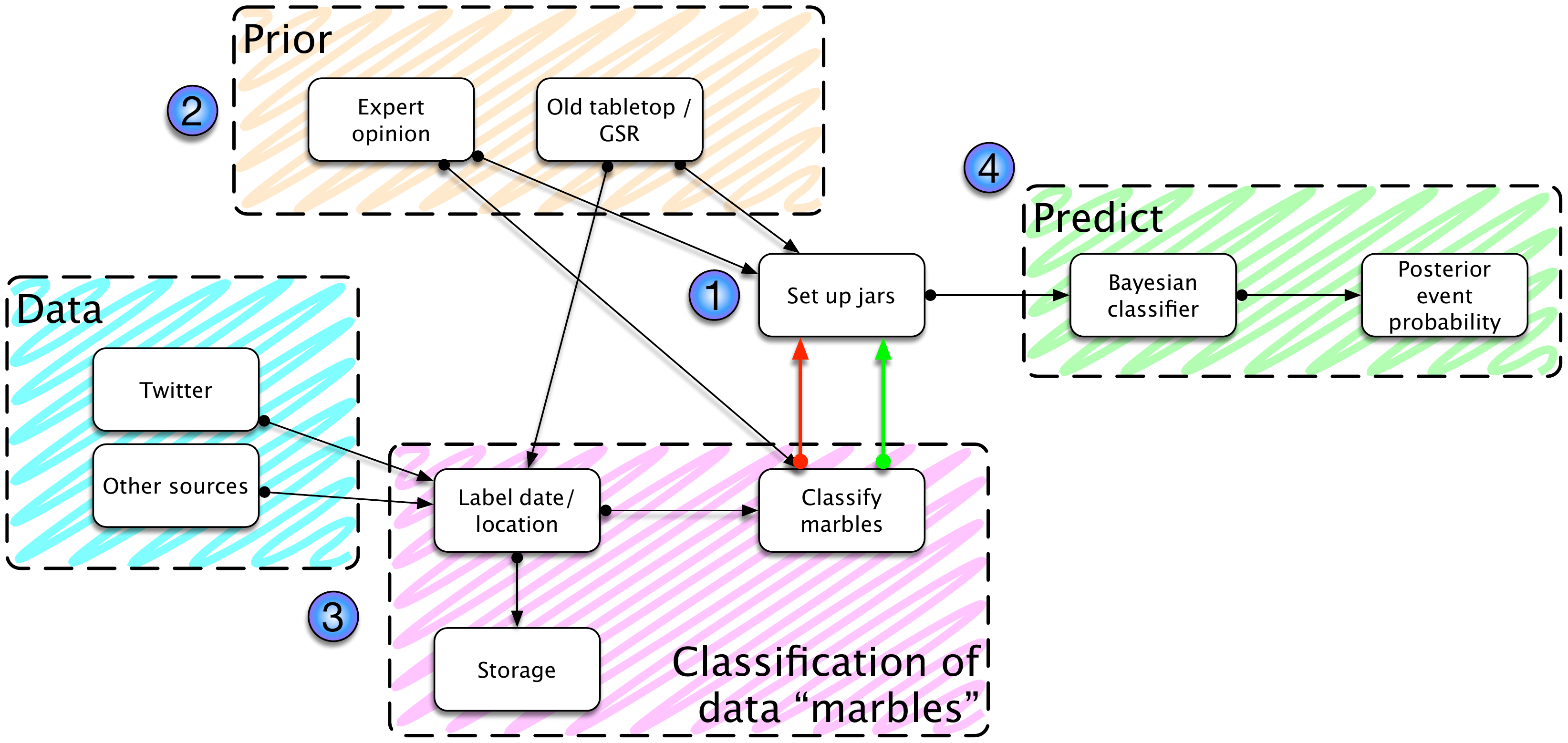}
\caption{Overview of the Pachinko Prediction (PP) framework.}
\label{fig:overview}
\end{figure}

An overview of our method is given in Figure~\ref{fig:overview}. 
This shows the stages of the prediction process: setting up appropriate data structures, then classifying individual social media postings as relevant for prediction or not, then making event predictions using a Bayesian classifier. 
The analogy we use throughout is that of a \emph{Bean machine}\footnote{No relation to N.~G.~Bean, an author of this paper.}, alternately called a \emph{Galton Board} or \emph{quincunx}, which illustrates the central limit theorem by using a table of pegs to sort marbles into jars. 
Our method can be conceptualised as a pegboard arrangement (an algorithm) to sort marbles (tweets) into jars (day-location pairs) for the purposes of making predictions.
These marbles can be either red or green (classified as being indicative of an event or not),
and the method works by monitoring the ratio of red to green marbles.
We will therefore refer to tweets as ``marbles'' when describing our method in this paper;
this analogy was a useful device for communicating our methodology to prospective end-users of this tool.
Historically the Bean machine is a precursor to the Japanese gambling game \emph{Pachinko}, where a large number of marbles are randomly sorted into bins via a pegboard -- some bins worth prizes, but most having no value. 
With our method also requiring to filter out a large volume of off-topic tweets in order to predict events,
we therefore refer to it as \emph{Pachinko Prediction} (PP).

The data structure for the method conceptualises a large table filled with a grid of jars, into which marbles will be sorted. 
First we label each jar with a date and a location. 
For example, we need jars for each of

\begin{itemize}
\item Melbourne, 21 July 2017, 
\item Melbourne, 22 July 2017, 
\item $\ldots$
\item Melbourne, 14 February 2018, 
\item Adelaide, 21 July 2017, 
\item $\ldots$
\end{itemize}

Each of these jars represent a date-location combination that is of interest to the researcher.

\subsection{Generative model}

\begin{figure}
\centering
\includegraphics[width = 0.5\textwidth]{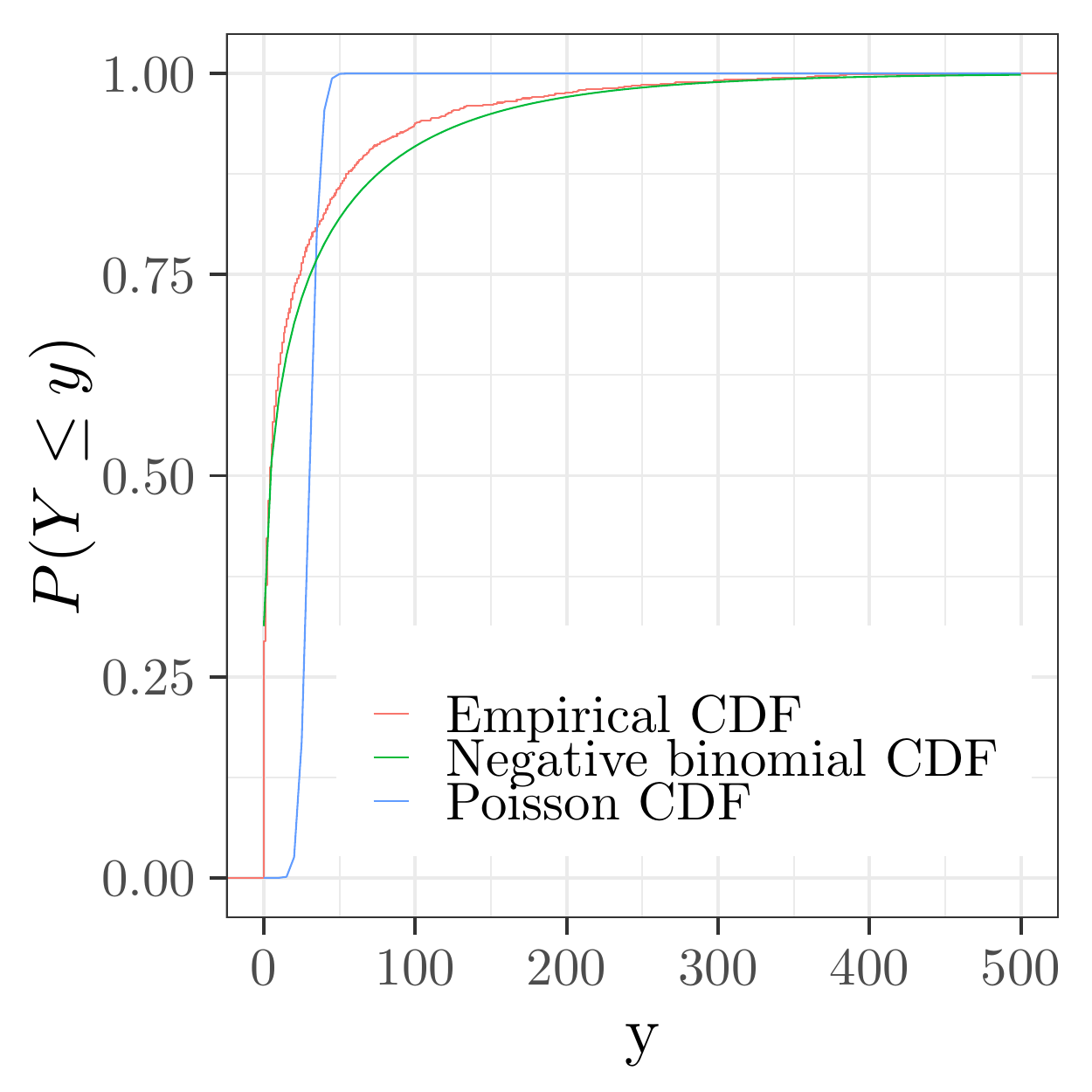}
\caption{Empirical cumulative distribution function for the number of tweets on a given day.}
\label{fig:marbles_cdf}
\end{figure}

For each day $i$ we counted the number of tweets mentioning that day. 
We denote the number of tweets for day $i$ as $Y_i$. 
Figure~\ref{fig:marbles_cdf}  plots the empirical cumulative distribution function (CDF) of $Y_i$. 
We considered two possible models to describe the number of tweets per day. 
The first was a Poisson distribution, \emph{i.e.}, 

$$
P(Y_i = y_i) = \frac{e^{-\lambda}\lambda^{y_i}}{y_i!}.
$$

We estimated the parameter $\lambda$ using maximum likelihood estimation with the {\tt fitdistplus} package \citep{Delignette-Muller:2015aa} in R, and obtained an estimate of $\hat{\lambda} = 30.82.$
The fitted CDF is given in Figure~\ref{fig:marbles_cdf}, and shows a large deviation away from the empirical CDF. 
This is confirmed if we examine the observed mean and variance of $Y_i$. 
The mean is 30.82, while the variance is 10199.21, indicating severe overdispersion. 

To deal with overdispersion for this Poisson random variable, we used the negative-binomial distribution, with parameters $\mu$, the mean number of observations, and $r$, used to deal with the overdispersion. 
There exist many different notations for the negative binomial distribution in the literature,
so to be specific, we define the negative binomial probability mass function here as 

$$
P(Y_i = y_i) = 
\frac{\Gamma(r + k)}{k! \Gamma(r)}
\left(
\frac{\mu}{r + \mu}
\right)^k
\left(
\frac{r}{r + \mu}
\right)^r.
$$

For instance, the version of the negative binomial in R switches $\mu$ and $(1 - \mu)$ in the above expression. 

Using maximum likelihood estimation, we obtain the following parameter estimates for the Twitter dataset: $\hat{\mu} = 30.67, \hat{r} = 0.24.$
Figure~\ref{fig:marbles_cdf} shows the CDF for the negative-binomial with these parameters, showing a far better correspondence to the empirical CDF.

\subsection{Bayesian classifier}

Based on our observations in Section \ref{sec:GSR} on associations between various different predictors we consider stratifying the observed data to best utilise this information for prediction.
We denote the probability of an event on day $i$ in location $j$ by $\theta_{k,ij}$, where $k$ is a stratification of the data on day $i$ in location $j$. 
In Section \ref{sec:results} we consider a number of different stratifications on the observed data.
We denote the number of red marbles (tweets indicative of a future event) mapped to day $i$ and location $j$, which are contained in strata $k$, as $Y_{k, ij}$.
Henceforth we will model only red marbles or \emph{indicative} tweets.
We found that the number of green marbles or non-indicative tweets were roughly constant across each day-location, making them non-informative for our analysis.
However, we note that the following method generalises trivially to utilising green marbles as well.


We assume that all the probabilities for days and cities contained in strata $k$ have the same prior distribution, which we denote $f(\theta_{k, \bullet\bullet})$. 
This assigns the same prior probability of an event on any given day in strata $k$, regardless of location.
While this prior could be customised for different cities (for example, it it reasonable to assume that larger cities have a higher probability of seeing events occur than smaller ones) we wanted to begin with a fairly weak prior.
We assume that this prior has a beta distribution with hyperparameters $\alpha$ and $\beta$, \ie, 

$$
f(\theta_{k,\bullet\bullet}) = 
\frac{\Gamma(\alpha)\Gamma(\beta)}{\Gamma(\alpha + \beta)}
\theta_{k, \bullet\bullet}^{\alpha - 1}
(1 - \theta_{k, \bullet\bullet})^{\beta - 1}.
$$


Let the number of indicative tweets in strata $k$ be $N_k$. 
Of these, $E_k$ occur on days having events.
We assume that $E_k$ given $\theta_{k,\bullet\bullet}$ is binomially distributed, \ie, 

$$
E_k|\theta_{k,\bullet\bullet} \sim Bin(N_k, \theta_{k, \bullet\bullet}), 
$$

which has the probability mass function

$$
f(E_k|\theta_{k, \bullet\bullet}) = \binom{N_k}{E_k} \theta_{k, \bullet\bullet}^{E_k} (1 - \theta_{k, \bullet\bullet})^{N_k - E_k}
$$

It can easily be shown, that the posterior distribution of $\theta_{k, \bullet\bullet}$, given $E_k$, is 

$$
f(\theta_{k, \bullet\bullet}  | E_K) 
\propto 
\theta_{k, \bullet\bullet}^{E_k + \alpha - 1}
(1 - \theta_{k, \bullet\bullet})^{N_k - E_k + \beta - 1}, 
$$
\ie, a beta distribution with parameters $E_k + \alpha$ and $N_k - E_k + \beta$. 


To estimate the hyperparameters $\alpha$ and $\beta$ we use an empirical Bayes approach and borrow information from the rest of the GSR dataset. 
We set $\alpha$ to be the number of event days across the entire country in the GSR dataset, 
and $\beta$ to be the number of non-event days in the GSR dataset. 
Note that we would obtain an equivalent parameterization if we used a non-informative prior and then obtained a posterior distribution for the overall proportion of an event. 


Now, we consider $Y_{k,ij}$, the number of indicative tweets from day $i$ in location $j$ which are contained in strata $k$. 
Based on the fitting we did in Figure \ref{fig:marbles_cdf}, 
we model the relationship between $Y_{k,ij}$ and $\theta_{k,ij}$ using a negative binomial model:

$$
f(Y_{k,ij}|\theta_{k,ij}) = 
\binom{y_{k,ij} + r - 1}{y_{k,ij}} \theta_{k,ij}^{y_{k,ij}} (1 - \theta_{k,ij})^{r}.
$$


Using the posterior distribution for $\theta_{k,\bullet\bullet}$ given $E_k$ as the prior, we get

$$
f(\theta_{k,ij} | Y_{k,ij}) 
\propto 
\theta_{k,ij}^{Y_{k,ij} + E_k + \alpha - 1}(1 - \theta_{k,ij})^{r + N_k - E_k + \beta - 1},
$$

which is once again a beta distribution. Note that we still need to estimate the parameter $r$. Once again we take an empirical Bayes approach, and use the MLE of $r$ from all indicative tweets. 

\section{Results}
\label{sec:results}

\subsection{Bayesian analysis}

In the previous section we outlined the Bayesian model that we will use to find the posterior distribution for an event on day $i$ in location $j$. 
In this model, we consider stratification of the observations to account for the observed relationship between location and month on the probability of an event in the GSR record (Section~\ref{sec:GSR}). 

We considered three possible strata, based on:
\begin{enumerate}
\item location,
\item month, and 
\item both location and month. 
\end{enumerate}
Tables~\ref{tab:city_strata} and \ref{tab:month_strata}
give the observed number of events and non-events for the location and month strata respectively. 
We omit the table of location+month strata for brevity.

\begin{table}[htbp]
\centering
\begin{tabular}{l|rrr}
  \hline
Location & Days & Events & No events \\ 
  \hline
Adelaide & 206 &  12 & 194 \\ 
  Brisbane & 209 &  37 & 172 \\ 
  Canberra & 206 &  27 & 179 \\ 
  Darwin & 208 &   5 & 203 \\ 
  Hobart & 207 &  11 & 196 \\ 
  Melbourne & 209 &  57 & 152 \\ 
  Perth & 209 &  30 & 179 \\ 
  Sydney & 209 &  47 & 162 \\ 
   \hline
\end{tabular}
\caption{Observed number of events and 
    non-events for each location strata.} 
\label{tab:city_strata}
\end{table}

\begin{table}[htbp]
\centering
\begin{tabular}{l|rrr}
  \hline
Month & Days & Events & No events \\ 
  \hline
Jul &  79 &  14 &  65 \\ 
  Aug & 248 &  41 & 207 \\ 
  Sep & 240 &  41 & 199 \\ 
  Oct & 248 &  40 & 208 \\ 
  Nov & 240 &  47 & 193 \\ 
  Dec & 248 &  20 & 228 \\ 
  Jan & 248 &  18 & 230 \\ 
  Feb & 112 &   5 & 107 \\ 
   \hline
\end{tabular}
\caption{Observed number of events and 
    non-events for each month strata.} 
\label{tab:month_strata}
\end{table}

\begin{figure}[htbp]
	\centering
	\includegraphics[width = 0.75\textwidth]{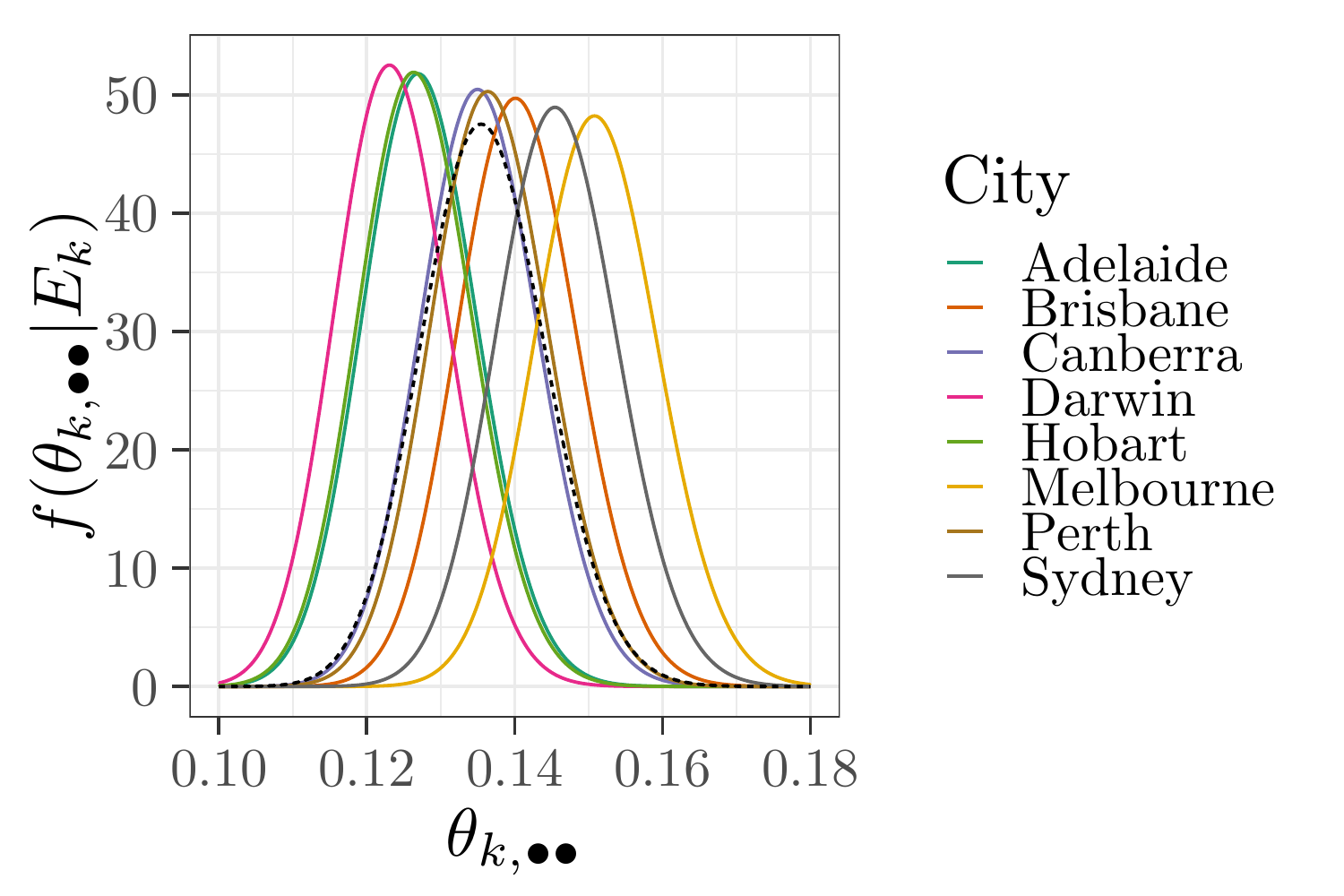}
	\caption{Posterior distribution of $\theta_{k,\bullet\bullet}$ given the city strata. The dashed line shows the overall posterior from all cities combined.}
	\label{fig:city_betas}
\end{figure}

Figure~\ref{fig:city_betas} shows the posterior distribution for $\theta_{k,\bullet, \bullet}$ given the location strata. 
For comparison, we have included the prior distribution for $\theta_{k,\bullet, \bullet}$ given only the overall number of event and non-events. 
For cities with an increased proportion of events, e.g., Melbourne, Sydney, and Brisbane, the distribution is shifted to the right, while for cities like Darwin, that have a low proportion of events, the posterior distribution is shifted to the left.
This is intuitive, as we observed in Section~\ref{sec:GSR} that larger cities are more likely to see more protest events.

\begin{figure}[htbp]
	\centering
	\includegraphics[width = \textwidth]{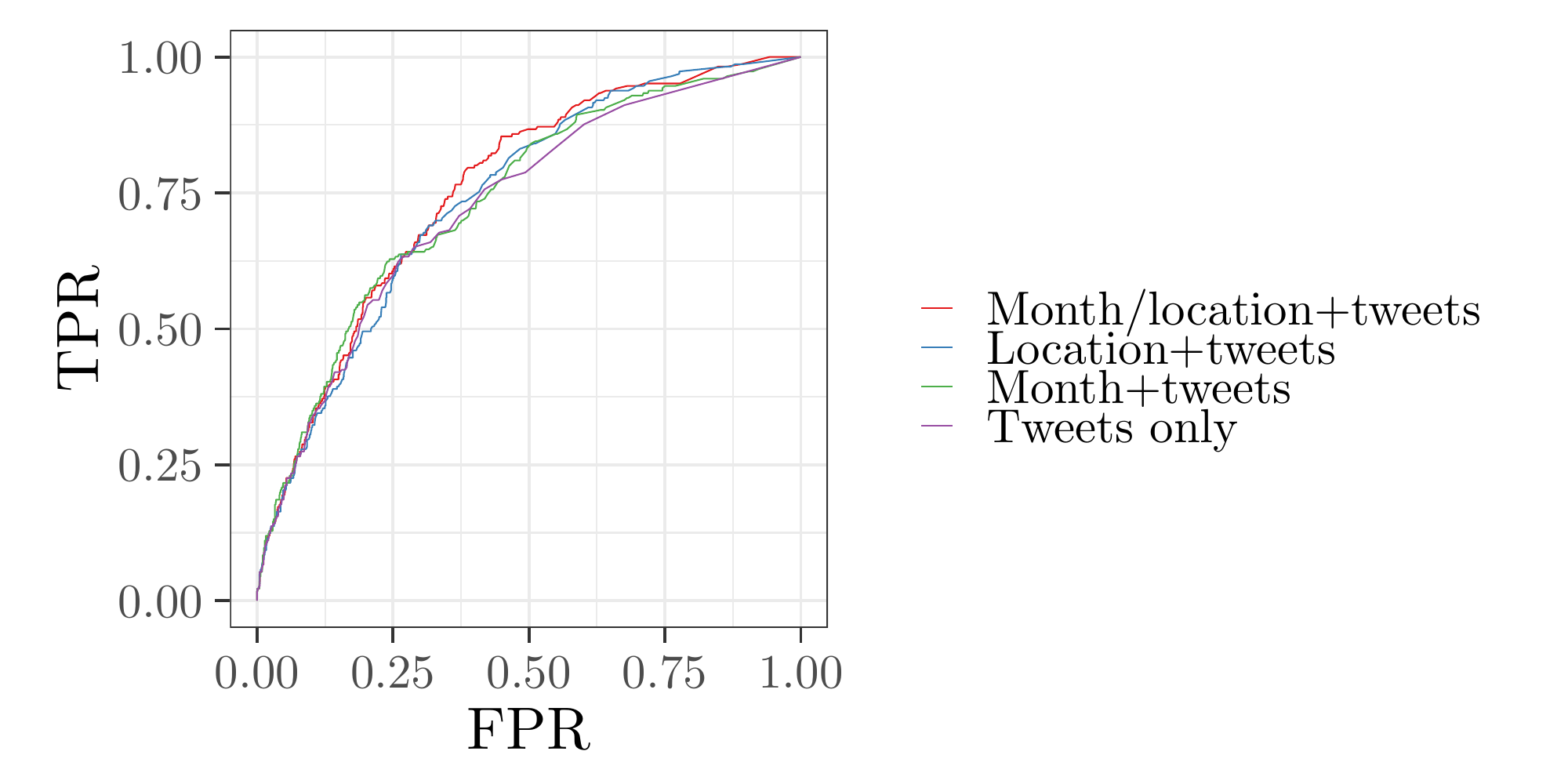}
	\caption{ROC curves for the different models using different stratifications.
	The \emph{Tweets only} model produces produces most of the gains over a baseline model,
	with the various stratifications adding further improvements to the ROC curves. Note that the Month/city strata without Twitter data performs comparably to the tweets-only model.}
	\label{fig:roc}
\end{figure}

To analyse the performance of our models, we choose to use Receiver Operating Characteristic (ROC) curves.
We do this because we do not wish to evaluate the model on a ``hard'' classification of predictions as being strictly event/non-event days.
Our model outputs a posterior distribution, from which we take the posterior mean as our prediction of the probability of seeing an event at day-location $ij$. 
To convert this probability to a binary prediction would require selecting an arbitrary threshold, above which we would issue an alert for an upcoming event.
On the other hand, the ROC curve allows us to visualise the performance of the model for \emph{all} possible thresholds, which respects the nature of the predictions output by the model better.
Figure~\ref{fig:roc} gives the ROC curves for a variety of models. We considered five models, as follows:


\begin{description}
	\item[Overall] This uses the overall number of events and non-events in the data and does not use any strata or any of the indicative tweets. This non-predictive model is equivalent to flipping a (biased) coin on each day we predict for. (Note: The ROC curve for this is by definition the line TPR = FPR, so we omit it from the figure for brevity.)
	\item[Tweets only] This uses just the overall number of  events and non-events in the data, and the observed number of indicative tweets for each day-location $ij$, but does not use any strata. It is effectively a ``data-only'' model, using a (close to) uninformative prior.
	\item[Location+tweets] This uses the strata based on the location, plus indicative tweet data.
	\item[Month+tweets] This uses the strata based on the month, plus indicative tweets. 
	\item[Month/location+tweets] This uses the strata based on the month and location, plus indicative tweets. 
\end{description}

We see an improvement in the ROC curves (contained in upper-left quadrant) once we start utilising the Twitter data. 
The best results are for the strata using the strata based on both the month and city. This is further seen in the Table~\ref{tab:roc} which gives the area under the ROC curve (AUC) for each model. 
We obtained similar AUC values using the other models.
Furthermore, performing the same experiment using cross validation with a random 70/30 train/test split produces comparable ROC curves showing similar AUC values.

\begin{table}[htbp]
\centering
\begin{tabular}{l|r}
  \hline
Strata & AUC \\ 
  \hline
Overall & 0.50 \\ 
  Tweets only & 0.73 \\ 
  Month+tweets & 0.74 \\ 
  Location+tweets & 0.74 \\ 
  Month/location+tweets & 0.76 \\ 
   \hline
\end{tabular}
\caption{Observed AUC for each of the models.} 
\label{tab:roc}
\end{table}

\begin{figure}[htbp]
	\centering
	\includegraphics[width = \textwidth]{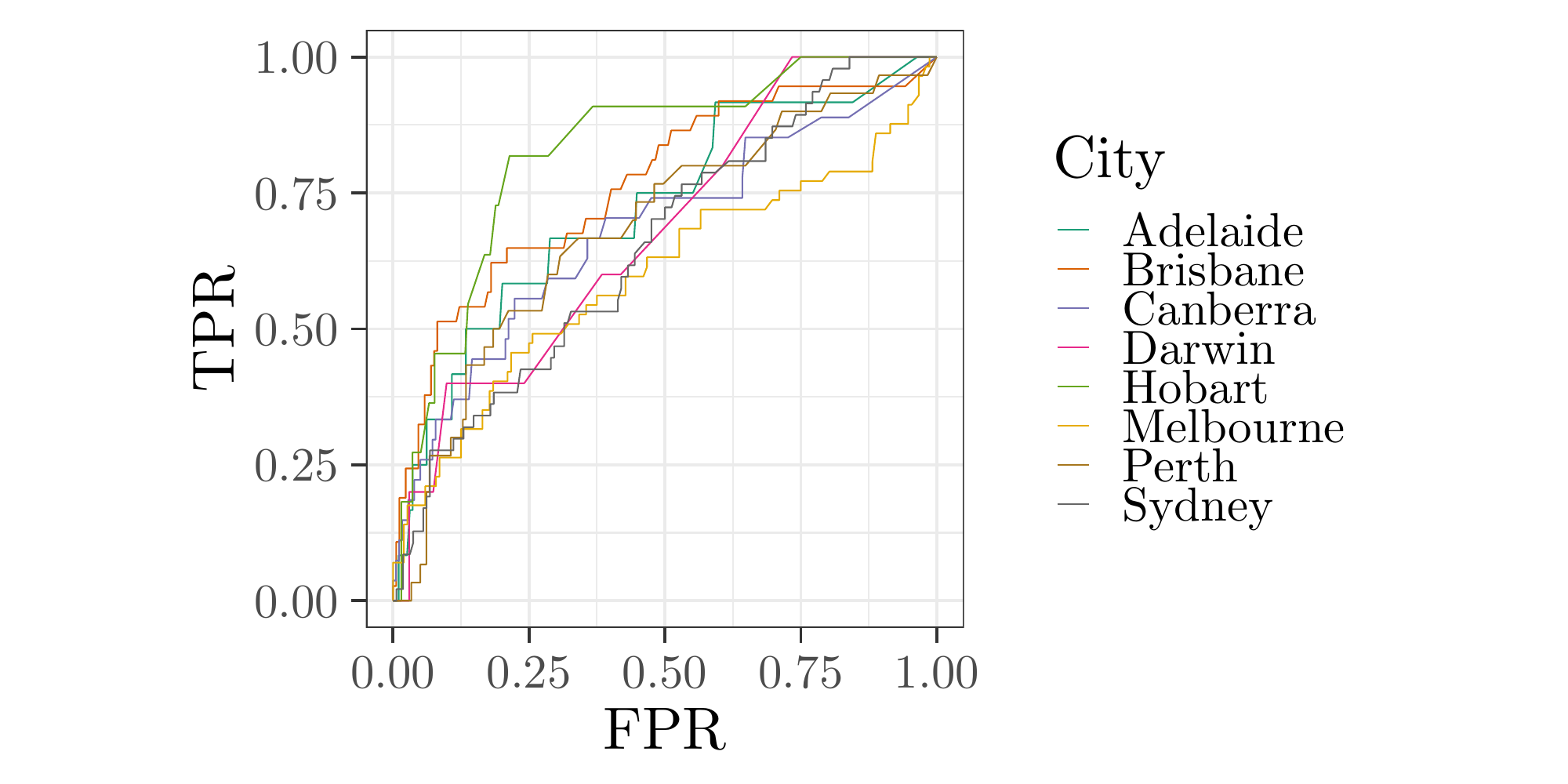}
	\caption{ROC curves for the different cities using the month/location+tweets model.}
	\label{fig:city_roc}
\end{figure}

We also considered the observed data by city to examine how well the model performs at predicting events in individual cities, rather than averaged Australia-wide. 
The ROC curves are given in Figure~\ref{fig:city_roc} and the AUC are given in Table~\ref{tab:city_roc}. 
We see that for Hobart the model performs particularly well, while the predictions for Melbourne are relatively poor, with an AUC of just 0.6. We explore this discrepancy further in the next subsection.
Once again, we obtain similar results when performing cross validation with a random 70/30 split (omitted for brevity).

\begin{table}[htbp]
\centering
\begin{tabular}{l|r}
  \hline
City & AUC \\ 
  \hline
Melbourne & 0.60 \\ 
  Sydney & 0.65 \\ 
  Canberra & 0.67 \\ 
  Darwin & 0.68 \\ 
  Perth & 0.68 \\ 
  Adelaide & 0.72 \\ 
  Brisbane & 0.76 \\ 
  Hobart & 0.83 \\ 
   \hline
\end{tabular}
\caption{Observed AUC for each city, using the month/location+tweets model.} 
\label{tab:city_roc}
\end{table}

\subsection{Prediction model}

\begin{sidewaysfigure}[htbp]
	\centering
	\includegraphics[width = 0.8\textwidth]{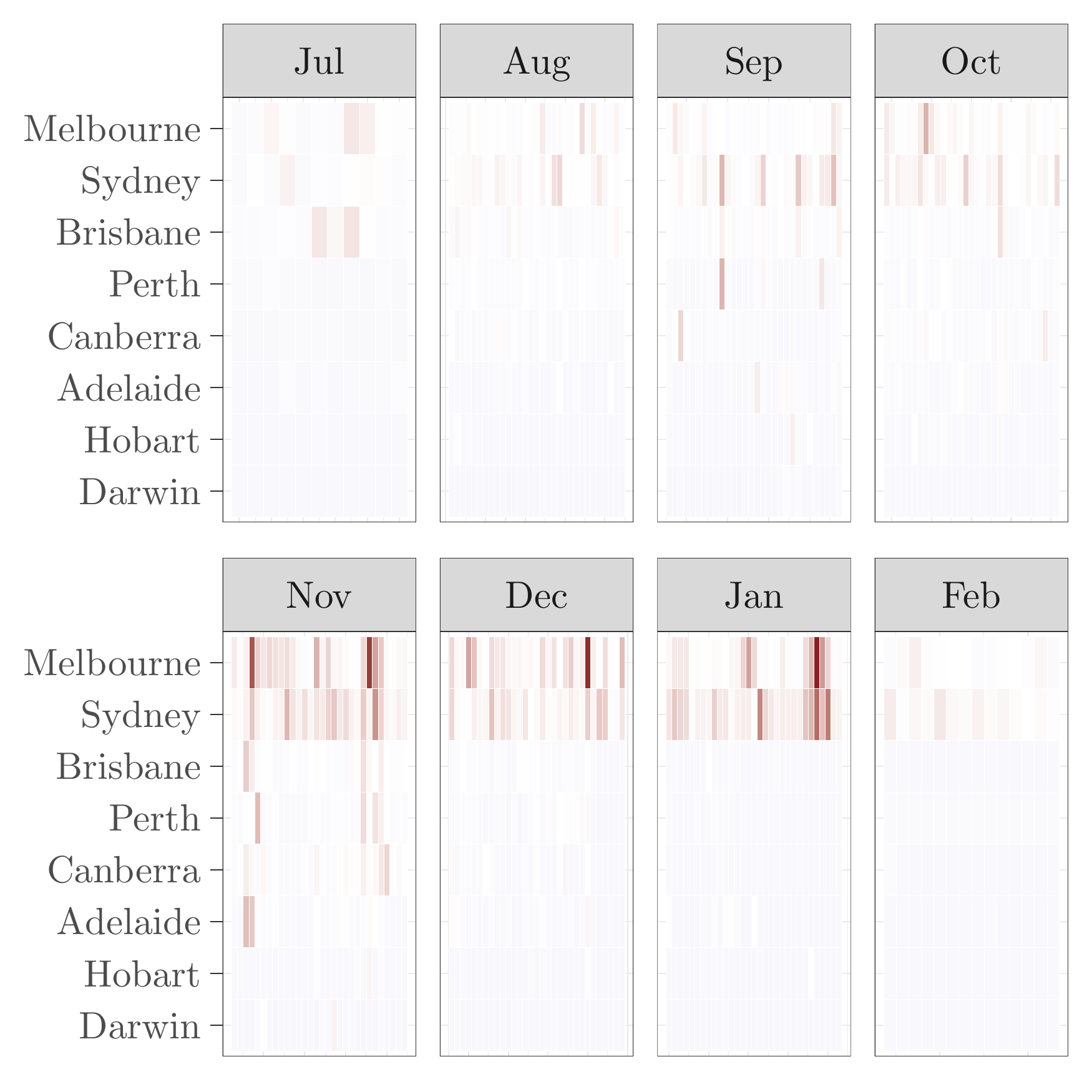}
	\caption{Time plot of predictions. Darker red indicates a higher prediction probability of an event.}
	\label{fig:time_plot_pred}
\end{sidewaysfigure}

Figure \ref{fig:time_plot_pred} shows predictions from our model over the full time period considered.
Comparison with Figure \ref{fig:time_plot} shows that the general trends of predictions are similar to those in the GSR,
with the model predicting higher probabilities of events occurring in larger cities Sydney and Melbourne.
More events were predicted to occur in the later months due to an increase in the volume of indicative tweets over this period.

\begin{figure}[htbp]
	\centering
	\includegraphics[width = 0.5\textwidth]{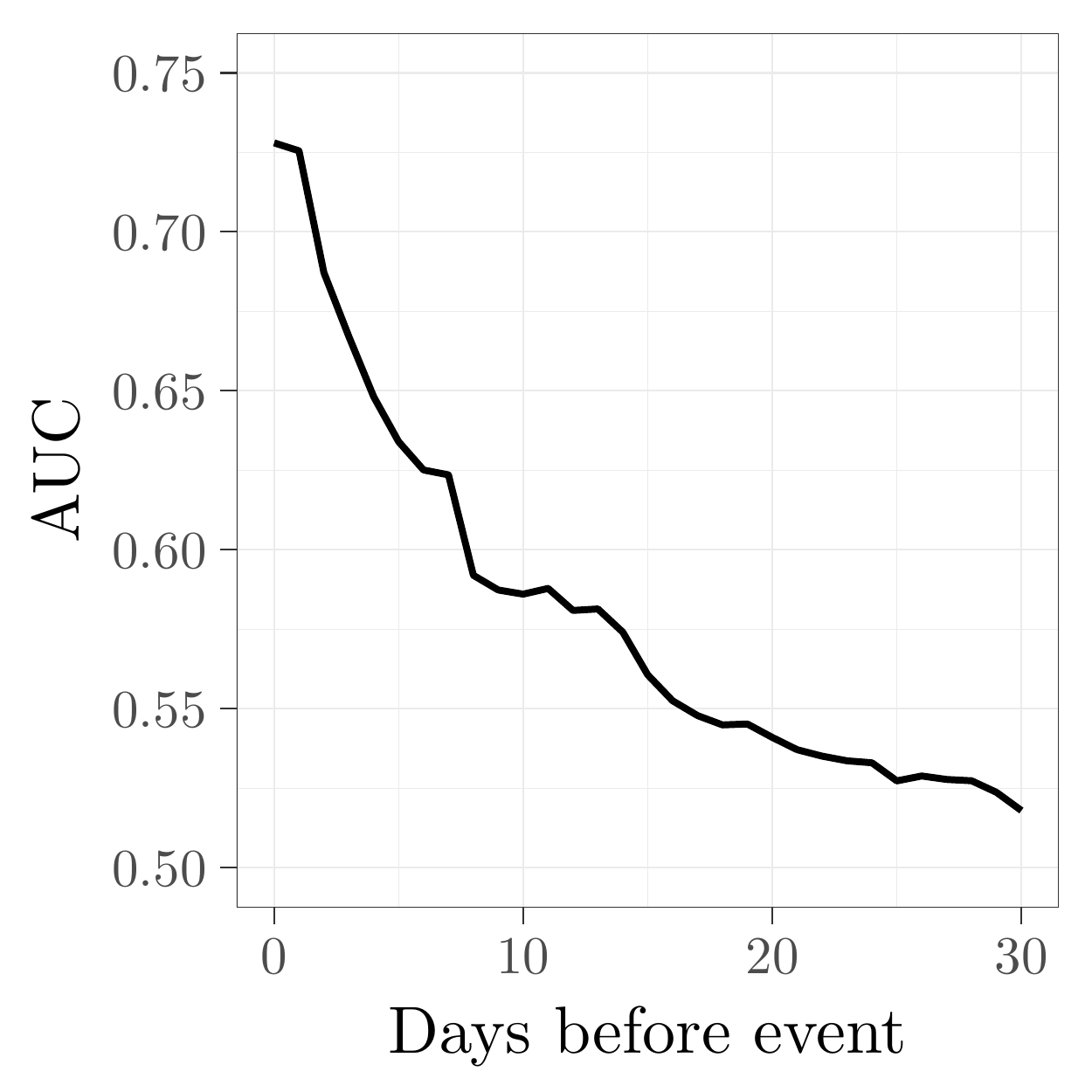}
	\caption{Decay in AUC for $n$-days-ahead prediction. There is a jump down at 7 days, but overall the model performs better than an uninformative model for up to 30 days.}
	\label{fig:roc_pred}
\end{figure}

To examine the amount of lead time in predictions made by our method, we performed $n$-days-ahead predictions, by considering only the tweets collected referring to an event that were authored up to $n$ days before that event.
Figure \ref{fig:roc_pred} shows the decay in AUC for predictions made based on the Twitter data available 0 to 30 days before each event.
Here we consider all cities, and we use the model with no stratification.
The model performs reasonably well up to a lead time of one week, after which there is a drop and the AUC continues to decay.
In particular, note that 1-day head predictions perform almost as well (in terms of AUC) as 0-day ahead do.
This indicates that there is consistently usable information contained within Twitter data in the day before an event.
For a potential end-user interested in advance warning of upcoming events this represents an actionable output from our model.
We note that even after 30 days the model still outputs some usable predictions, performing slightly better than an uninformative (coin-flip) model, with an AUC slightly above 0.5.

\begin{figure}[htbp]
	\centering
	\includegraphics[width = 0.75\textwidth]{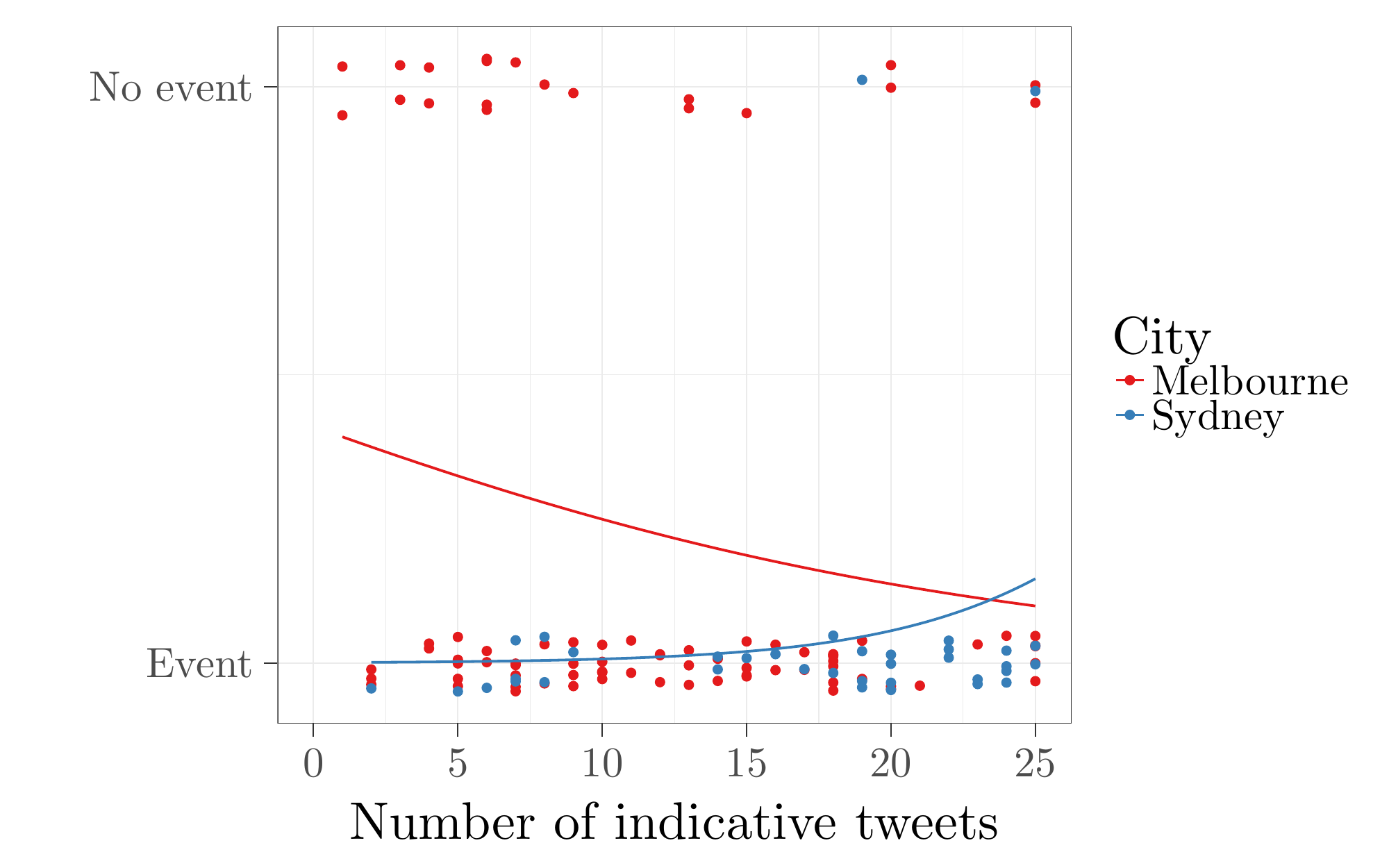}
	\caption{Events and non-events for the days with low numbers of indicative tweets for Sydney and Melbourne. The lines are fitted logistic regressions for each city. Models were fitted on all data; here we focus only on days having 25 indicative tweets or less. We also added vertical jitter to the data points for visibility.}
	\label{fig:low_RM}
\end{figure}

To examine why predictions for Melbourne are relatively poor relative to the other cities, we compared Sydney and Melbourne, being two cities of roughly equal population. They also had a similar number of events over the period: 57 for Melbourne, and 47 for Sydney.
The main difference between the cities becomes clear when we examine the days with low numbers of indicative tweets more closely.
Figure~\ref{fig:low_RM} shows days with 25 indicative tweets or less in Sydney and Melbourne. 
While for Sydney there are only two event days containing 25 indicative tweets or less, we observe that there were a large number of events occurring in Melbourne for which very few indicative tweets were detected by our system. 
The lines are logistic regression fits to the data for each city. 
These make clear that for days having a small number of indicative tweets, the logistic regression predicts a decreasing probability of an event for increasing number of indicative tweets for Melbourne. 
This suggests that either there were few tweets authored referring to these events, or that our protest classifier described in Section \ref{sec:datasets} is poorly tuned to detect these events.

Examining the events in Melbourne with fewer than 25 tweets shows that these events often concern smaller sub-populations. 
For example, some of the headlines for corresponding to these low-tweet events include:

\begin{quote}
    \texttt{Vic pathology staff on indefinite strike}
    
    \texttt{Accused Neil Erikson in bid to dismiss mock beheading charges}
    
    
    \texttt{DVA rally: Families want royal commission after series of veteran suicides}
    
    \texttt{Airport workers in Melbourne stage protest}
\end{quote}

In each of these cases, the sub-population involved is small (e.g. pathology staff, or families of veterans). 
These groups likely use different methods (potentially other social media platforms, or a different medium entirely) to organise these events.
Indeed, when we went back to the historical Twitter record to search for tweets mentioning these events before they occurred,
we found only a very small number of tweets.
It is clearly the case that PP will struggle to detect events concerning only small sub-populations of this type.
It has been observed previously for other methods (even EMBERS) that signals for all types of events do not necessarily appear in all types of data sources \citep{Korkmaz2016}.
We remark that while Twitter may not be the appropriate medium for detecting these particular events, it is likely that combining multiple datasets (e.g., Facebook posts or appropriate web searches) would improve our predictions, and our framework is flexible to allow for doing this in the same manner as we have done here for tweets.


\section{Discusssion}
\label{sec:discussion}

In this paper we have developed a Bayesian methodology for predicting events from Twitter data.
Our method makes the contribution of being interpretable,
with the ability to both explicitly show the evidence upon which a particular prediction is made,
as well as separating between the components of the prediction coming from the Twitter data and the prior belief about event probabilities.
Furthermore,
the model predicts the probability of an event occurring, rather than a binary classification of a particular day being an event/non-event day.
This empowers a greater understanding of the uncertainties associated with each prediction,
and gives the end-user an indication of how much confidence they should place in any given prediction.
Combined with the clear ``audit trail'' of evidence underlying each prediction output by our model,
we argue that this facilitates more informed decision-making by potential end-users.

The framework developed here naturally generalises to incorporating multiple heterogeneous data sources for predictions as in \cite{Korkmaz2016}.
Future work will explore methods to incorporate other open data sources such as Facebook pages or search activity.
We will also look at predicting other characteristics of the events contained within the GSR data such as population group,
type of event, and whether it is likely to be violent or non-violent.

In this work we treated individual tweets as being independent, which of course is unlikely to be a valid assumption.
With the authors of these tweets being embedded in a complex social network, there exist clear dependencies between their activity patterns \citep{Bagrow2017b} which are unaccounted for here.
Utilising this network structure between accounts may help uncover bot accounts \citep{Nasim2018}, which while potentially still useful for making predictions, should at the very least be accounted for in the model.
Future work will investigate using network characteristics of tweets referencing the same day as features in predictive models.
For example, networks of tweets may be formed with edges representing some shared characteristic between those tweets (e.g., common hashtags, authors, or replies).
Using structural characteristics of these networks may reduce ``noise'' in the data used for prediction, and produce better-quality evidence for upcoming events.

\bibliographystyle{apalike}
\bibliography{papers}

\begin{thebibliography}{}

\bibitem[Agarwal and Sureka, 2016]{Agarwal2016}
Agarwal, S. and Sureka, A. (2016).
\newblock {Investigating the Potential of Aggregated Tweets as Surrogate Data
  for Forecasting Civil Protests}.
\newblock In {\em Proceedings of the 3rd IKDD Conference on Data Science, 2016
  (CODS `16)}.

\bibitem[Alajajian et~al., 2017]{Alajajian2017}
Alajajian, S.~E., Williams, J.~R., Reagan, A.~J., Alajajian, S.~C., Frank,
  M.~R., Mitchell, L., Lahne, J., Danforth, C.~M., and Dodds, P.~S. (2017).
\newblock {The lexicocalorimeter: Gauging public health through caloric input
  and output on social media}.
\newblock {\em PLoS ONE}, 12(2):e0168893.

\bibitem[Alanyali et~al., 2015]{Alanyali2015}
Alanyali, M., Preis, T., and Moat, H.~S. (2015).
\newblock {Tracking Protests Using Geotagged Flickr Photographs.}
\newblock {\em Under Review}, pages 27--30.

\bibitem[Bagrow et~al., 2017]{Bagrow2017b}
Bagrow, J.~P., Liu, X., and Mitchell, L. (2017).
\newblock {Information flow reveals prediction limits in online social
  activity}.
\newblock {\em arXiv preprint}, 1708.04575.

\bibitem[Bollen et~al., 2011]{Bollen2011}
Bollen, J., Mao, H., and Zeng, X. (2011).
\newblock {Twitter mood predicts the stock market}.
\newblock {\em Journal of Computational Science}, 2(1):1--8.

\bibitem[Borge-Holthoefer et~al., 2016]{Borge-Holthoefer2015a}
Borge-Holthoefer, J., Perra, N., Gon{\c{c}}alves, B.,
  Gonz{\'{a}}lez-Bail{\'{o}}n, S., Arenas, A., Moreno, Y., and Vespignani, A.
  (2016).
\newblock {The dynamics of information-driven coordination phenomena: A
  transfer entropy analysis}.
\newblock {\em Science Advances}, 2(4).

\bibitem[Cadena et~al., 2015]{Cadena2015}
Cadena, J., Korkmaz, G., Kuhlman, C.~J., Marathe, A., Ramakrishnan, N., and
  Vullikanti, A. (2015).
\newblock {Forecasting Social Unrest Using Activity Cascades.}
\newblock {\em PLoS ONE}, 10(6):e0128879.

\bibitem[Chang and Manning, 2012]{Chang2012}
Chang, A.~X. and Manning, C.~D. (2012).
\newblock {SUTime: A library for recognizing and normalizing time expressions.}
\newblock {\em Proceedings of the Eighth International Conference on Language
  Resources and Evaluation (LREC-2012)}, (iii):3735--3740.

\bibitem[Cody et~al., 2015]{Cody2015}
Cody, E.~M., Reagan, A.~J., Mitchell, L., Dodds, P.~S., and Danforth, C.~M.
  (2015).
\newblock {Climate change sentiment on Twitter: An unsolicited public opinion
  poll}.
\newblock {\em PLoS ONE}, 10(8):e0136092.

\bibitem[Delignette-Muller et~al., 2015]{Delignette-Muller:2015aa}
Delignette-Muller, M.~L., Dutang, C., et~al. (2015).
\newblock fitdistrplus: An {R} package for fitting distributions.
\newblock {\em Journal of Statistical Software}, 64(4):1--34.

\bibitem[Gallagher et~al., 2018]{Gallagher2018}
Gallagher, R.~J., Reagan, A.~J., Danforth, C.~M., and Dodds, P.~S. (2018).
\newblock {Divergent discourse between protests and counter-protests:
  {\#}BlackLivesMatter and {\#}AllLivesMatter}.
\newblock {\em PLoS ONE}, 13(4):1--23.

\bibitem[Hoegh et~al., 2015]{Hoegh2015}
Hoegh, A., Leman, S., Saraf, P., and Ramakrishnan, N. (2015).
\newblock {Bayesian Model Fusion for Forecasting Civil Unrest}.
\newblock {\em Technometrics}, 57(3):332--40.

\bibitem[Korkmaz et~al., 2016]{Korkmaz2016}
Korkmaz, G., Cadena, J., Kuhlman, C.~J., Marathe, A., Vullikanti, A., and
  Ramakrishnan, N. (2016).
\newblock {Multi-source models for civil unrest forecasting}.
\newblock {\em Social Network Analysis and Mining}, 6(1).

\bibitem[Mitchell, 2018]{Mitchell2018a}
Mitchell, L. (2018).
\newblock {Civil unrest event-relevant Twitter classifier training data}.
\newblock {\em Mendeley Data, v2}, doi:10.17632/mxcsxp3jxn.2.

\bibitem[Muthiah et~al., 2015]{Muthiah2015}
Muthiah, S., Huang, B., Arredondo, J., Mares, D., Getoor, L., Katz, G., and
  Ramakrishnan, N. (2015).
\newblock {Planned protest modeling in news and social media}.
\newblock In {\em Proceedings of the National Conference on Artificial
  Intelligence}, volume~5, pages 3920--3927.

\bibitem[Nasim et~al., 2018]{Nasim2018}
Nasim, M., Nguyen, A., Lothian, N., Cope, R., and Mitchell, L. (2018).
\newblock {Real-time detection of content polluters in partially observable
  Twitter networks}.
\newblock In {\em Proceedings of the 26th International Conference on the World
  Wide Web (WWW '18) Companion}, pages 1331--1339.

\bibitem[Pedregosa et~al., 2011]{scikit-learn}
Pedregosa, F., Varoquaux, G., Gramfort, A., Michel, V., Thirion, B., Grisel,
  O., Blondel, M., Prettenhofer, P., Weiss, R., Dubourg, V., Vanderplas, J.,
  Passos, A., Cournapeau, D., Brucher, M., Perrot, M., and Duchesnay, E.
  (2011).
\newblock Scikit-learn: Machine learning in {P}ython.
\newblock {\em Journal of Machine Learning Research}, 12:2825--2830.

\bibitem[Ramakrishnan et~al., 2014]{Ramakrishnan2014}
Ramakrishnan, N., Butler, P., Muthiah, S., Self, N., Khandpur, R., Saraf, P.,
  Wang, W., Cadena, J., Vullikanti, A., Korkmaz, G., Kuhlman, C., Marathe, A.,
  Zhao, L., Hua, T., Chen, F., Lu, C.-t., Huang, B., Srinivasan, A., Trinh, K.,
  Getoor, L., Katz, G., Doyle, A., Ackermann, C., Zavorin, I., Ford, J.,
  Summers, K., Fayed, Y., Arredondo, J., Gupta, D., and Mares, D. (2014).
\newblock {`Beating the news' with EMBERS: Forecasting Civil Unrest using Open
  Source Indicators}.
\newblock In {\em Proceedings of the 20th ACM SIGKDD international conference
  on Knowledge discovery and data mining (KDD '14)}, pages 1799--1808.

\bibitem[Str\"{o}tgen and Gertz, 2013]{Stroetgen2013}
Str\"{o}tgen, J. and Gertz, M. (2013).
\newblock Multilingual and cross-domain temporal tagging.
\newblock {\em Language Resources and Evaluation}, 47(2):269--298.

\bibitem[Theocharis et~al., 2015]{Theocharis2015}
Theocharis, Y., Lowe, W., van Deth, J.~W., and Garc{\'{i}}a-Albacete, G.
  (2015).
\newblock {Using Twitter to mobilize protest action: online mobilization
  patterns and action repertoires in the Occupy Wall Street, Indignados, and
  Aganaktismenoi movements}.
\newblock {\em Information Communication and Society}.

\bibitem[Xu et~al., 2014]{Xu2014}
Xu, J., Lu, T.~C., Compton, R., and Allen, D. (2014).
\newblock {Civil unrest prediction: A Tumblr-based exploration}.
\newblock In {\em Lecture Notes in Computer Science (including subseries
  Lecture Notes in Artificial Intelligence and Lecture Notes in
  Bioinformatics)}, volume 8393 LNCS.

\end{thebibliography}

\end{document}